\documentclass[final,onecolumn,12pt]{IEEEtran}
\usepackage[T1]{fontenc}
\usepackage[english]{babel}
\usepackage{amssymb,amsmath}
\usepackage{amsfonts}
\usepackage{verbatim}
\usepackage{dsfont}
\usepackage{cite}
\usepackage{graphicx}
\usepackage{mathtools}
\usepackage{subfigure}
\usepackage{tikz}
\usetikzlibrary{arrows,automata,positioning}
\usepackage{pgfplots}
\pgfplotsset{compat=newest}

\newtheorem{defn}{Definition}

\begin{document}

\title{On The Age Of Information In Status Update Systems With Packet Management}
\author{\IEEEauthorblockN{Maice Costa, Student Memeber, IEEE, Marian Codreanu, Member, IEEE,\\ and Anthony Ephremides, Life Fellow, IEEE}
\thanks{M. Costa and A. Ephremides are with the Department of Electrical and Computer Engineering at the University of Maryland, College Park, MD, USA 20742. Email: \{mcosta,etony\}@umd.edu}
\thanks{Marian Codreanu is with the Centre for Wireless Communications, University of Oulu, Oulu, Finland 90570. Email: codreanu@ee.oulu.fi}     
\thanks{This work was partially supported by NSF Grants CCF-0905209 and CCF-1420651, by ONR Grant N000141410107, and by Academy of Finland. }
\thanks{This work was presented in part at the IEEE International Symposium on Information Theory, ISIT 2014. }    
} 

\maketitle

\begin{abstract}

We consider a communication system in which status updates arrive at a source node, and should be transmitted through a network to the intended destination node. The status updates are samples of a random process under observation, transmitted as packets, which also contain the time stamp to identify when the sample was generated. The \textit{age} of the information available to the destination node is the time elapsed since the last received update was generated. In this paper, we model the source-destination link using queuing theory, and we assume that the time it takes to successfully transmit a packet to the destination is an exponentially distributed service time. We analyze the \textit{age of information} in the case that the source node has the capability to manage the arriving samples, possibly discarding packets in order to avoid wasting network resources with the transmission of stale information. In addition to characterizing the \textit{average age}, we propose a new metric, called \textit{peak age}, which provides information about the maximum value of the \textit{age}, achieved immediately before receiving an update. 
\end{abstract}


\section{Introduction}\label{sec:Intro}
Numerous applications of communication networks require the transmission of information about the state of a process of interest between a source and a destination. This is the case, for example, in sensor networks, where sensor nodes report the observations to a central processor, to monitor health or environment conditions \cite{Ko2010,Corke2010}. Another example is the use of a feedback channel to report side information in communication networks \cite{Ekpenyong2007, Ying2011}. In these applications, the timeliness of the transmitted message is an important and often critical objective, since an outdated message may lose its value. Hence, a theory of \textit{age of information} is useful to optimize communication systems when the receiver has interest in fresh information. This work is a contribution to the initial efforts in this direction.

We focus on applications of communication networks in which a random process is observed, and samples are made available to a source node at random times. The samples are transmitted in order to update the value of the process known at the destination node. For that reason, the transmitted messages are also called status updates. 

Status updates are transmitted as packets, containing information about one or more variables of interest, and the time of generation of the sample. Typically, it takes a random time until the packet is successfully delivered through the network. If the most recently received update carries the time stamp $U(t)$, which characterizes the instant of generation of the update, then the \textit{status update age}, or simply \textit{age}, is defined as the random process $\Delta(t)=t-U(t)$ \cite{Kaul2012}. Hence, the \textit{age} is the time elapsed since the last received packet was generated.

In this work, we model the source-destination link as a single-server queue, and consider a particular aspect in the transmission of status updates, namely the \textit{packet management} at the source node. We assume that the source receives random updates, but some of the packets may be discarded even before being transmitted through the network. Part of the results have been presented in \cite{Costa2014}.

We consider three policies for packet management: \textbf{(\textit{i})} discard every packet that finds the server busy,  \textbf{(\textit{ii})} keep a single packet waiting for transmission, and discard any additional packets that find the system full upon arrival, and \textbf{(\textit{iii})} keep a single packet waiting for transmission, replacing it upon arrival of a more up-to-date packet. 

The \textit{average age} is calculated for each of the three policies, and we compare the results with a scheme without packet management. One remark is that the packet management policies are adequate if the objective is to inform the destination about the most recent value of the process of interest, as we do not consider any cost associated with discarding samples. There may be applications where the availability of multiple consecutive samples is also desirable, even at the cost of a larger associated \textit{age}, for example, if an estimation is performed at the destination node. This interesting trade-off is out of the scope of this paper. 

In addition to the analysis of the \textit{average age}, we propose a new metric, called \textit{peak age}, which provides information about the maximum value of \textit{age} achieved immediately prior the reception of an update. The \textit{peak age} can be used to characterize the timeliness of information transmitted through a network, instead of the \textit{average age}, with the advantage of a more simple formulation, and a complete description of its probability distribution. Another motivation to define the \textit{peak age} is its relationship with the probability that the \textit{age} exceeds a threshold, which is relevant to design systems with a guarantee that the information available at the destination is fresh at any given time.

\subsection{Related Work}\label{sec:literature}

The \textit{age of information} was formalized as a metric of interest in the context of vehicular ad-hoc networks in \cite{Kaul2011} and \cite{Kaul2011a}. In \cite{Kaul2011}, the authors address the problem of congestion control in large vehicular networks, proposing a rate control algorithm to minimize the \textit{age of information} throughout the system. The effect of piggybacking messages throughout the vehicular network was investigated in \cite{Kaul2011a}, and shown to be effective in reducing the \textit{age of information}.

Simple network models based on queuing theory have been used in \cite{Kaul2012}, \cite{Yates2012}, and \cite{Kaul2012a}, aiming to understand more fundamental characteristics of the process describing the \textit{age of information}. The \textit{average age} has been investigated in \cite{Kaul2012}, considering randomly generated samples arriving according to a Poisson process, transmitted between a single source-destination pair, using a first-come-first-served (FCFS) discipline. With the simplifying assumption that the time it takes for a sample to be transmitted is exponentially distributed, and assuming that all samples wait in a queue for transmission, the authors model the system as an M/M/1 queue. The \textit{average age} for the M/D/1 and D/M/1 models is also calculated in \cite{Kaul2012}, illustrating the cases with deterministic service time and periodic sampling, respectively. The case with multiple sources was studied in \cite{Yates2012}. The main result is that optimizing the system for timely updates is not the same as maximizing throughput, nor it is the same as minimizing delay. The work in \cite{Kaul2012a} is more closely related to ours. The authors have analyzed the case with last-come-first-served (LCFS) queue discipline, and have shown that this transmission discipline achieves a lower bound for the \textit{average age} when the arrival rate is very large. In this work, we show that the same lower bound can be achieved with first-come-first-served (FCFS) discipline and packet management policies.

The transmission of status updates through a network cloud was investigated in \cite{Kam2013}, where the authors calculate the \textit{average age}, considering that all packets are transmitted immediately after generation, and some packets are rendered obsolete, due to the random service times in the network. This model corresponds to the extreme case of an infinite number of servers, and it presents a case in which resources are wasted transmitting all the available samples. In \cite{Kam2014}, the authors discuss the case with only two servers, accounting for both the effect of queuing and the possibility that packets are received out of order, wasting network resources. The packet management schemes proposed in this work prevent the transmission of outdated messages and the consequential waste of network resources. 

\subsection{Organization}
The formal definition of \textit{age of information}, the proposed packet management policies, and the corresponding system models are described in section \ref{sec:model}. In section \ref{sec:metrics}, we describe the two metrics for \textit{age of information} investigated in this paper, namely the \textit{average age} and the \textit{peak age}. The analytic results for \textit{average age} and \textit{peak age} are discussed in section \ref{sec:analytic}, in subsections dedicated to each of the three packet management schemes. Numerical results are presented in section \ref{sec:numerical}, and final remarks in section \ref{sec:conclusion}.

\section{Problem Statement and System Model}\label{sec:model}

Consider a communication link with one source-destination pair. A random process $H(t)$ is observed, and the status of this process is available to the source node at random time instants. The destination node has interest in timely information about the status of the process $H(t)$, but this information is not available instantaneously, since it has to be transmitted from the source through a link with limited resources. The source node obtains samples of the random process of interest $H(t)$, and transmits status updates in the form of packets, containing the information about the status of the process, and the time instant that the sample was generated. We denote the time stamps with $t_k$, and each status update message contains the information $\{H(t_k),t_k\}$. 

\begin{figure}
\centering
\begin{tikzpicture}
\draw[->,thick](-1.2,0)--(2.5,0)node[below, pos=0.95]{t};
\draw[->,thick](-1,-0.2)--(-1,3.5) node[pos=0.99,right]{Process of interest H(t)};
\draw[dashed, blue,thick](0.5,2)--(0.5,0) node [below] {$t_1$};
\draw[dashed,red,thick](1,0.5)--(1,0) node [below] {$t_2$};
\draw [black,thick] plot [smooth, tension=1] coordinates { (-1,1) (-0.5,3) (0,1) (0.5,2) (1,0.5) (1.5,2.2) (2,0.3)};
\draw[->,very thick](2.5,1.5)--(4.5,1.5) node [midway,below, font=\fontsize{20}{20}\selectfont]{$\lambda$}; 
\node at (3.5,1.9) [red,thick]{$\{H(t_2),t_2\}$};
\draw (4.8,2)--(7,2)--(7,1)--(4.8,1) node [below] at (5.9,1) {Buffer};
\node at (5.9,1.5) [shape=rectangle, blue, thick]{$\{H(t_1),t_1\}$};
\draw[->,very thick](7,1.5)--(8,1.5);
\draw[thick](8.5,1.5) node [font=\fontsize{20}{20}\selectfont]{$\mu$} circle (0.5); 
\draw[->,very thick](9,1.5)--(10,1.5);
\draw (10,1) rectangle(11,2)  node [font=\fontsize{20}{20}\selectfont,pos=0.5]{$\Delta$};
\end{tikzpicture}
\caption{Transmission of status updates through queuing system.}
\label{fig:UpdatesQueue}
\end{figure}
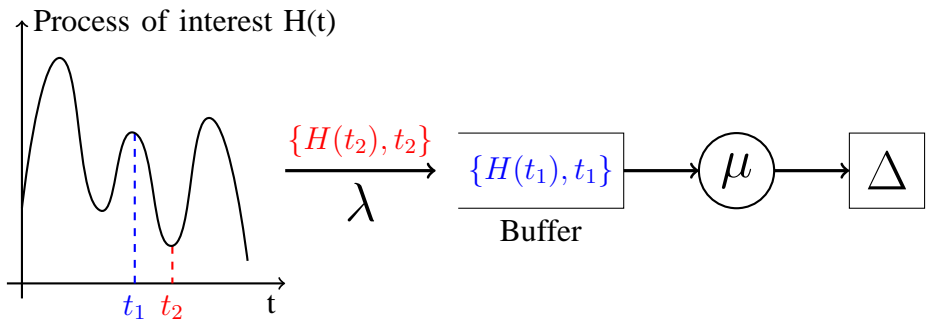

The transmission of status update messages will be studied using queuing theory, and we illustrate the system in Figure \ref{fig:UpdatesQueue}. We represent the random process of interest $H(t)$, which is observed in different time instants $t_k$, $k=1,2,\ldots$. Each observation becomes immediately available to the source node as a packet containing $\{H(t_k),t_k\}$. Hence, $t_k$ is also regarded as the arrival time of the packet to the source node. The arrival process is modeled as a Poisson process of rate $\lambda$, and each packet may be stored in a finite capacity buffer, if space is available, or discarded according to the packet management scheme. The transmission of a packet takes a random amount of time, which depends on channel conditions such as fading, and network conditions such as congestion. As a simplifying assumption, we consider that the time for transmission of a packet is exponentially distributed, with mean $1/\mu$.  The adopted queue models have a single server, and packets are transmitted using a first-come-first-served (FCFS) policy. In the destination, we are interested in characterizing the timeliness of the information available about the process $H(t)$.

The timeliness of the information about the process of interest can be characterized by the \textit{age of information} \cite{Kaul2012}, which we define below.

\begin{defn}{Age of Information}

Let $H(t)$ be a random process of interest. Let status update messages be packets containing the information $\{H(t_k),t_k\}$, $k=1,2,\ldots$, to be transmitted in a source-destination link. Denote with $t{'}_{k}$ the corresponding time instants at which the status update messages arrive at the destination. 

At a time instant $s$, we define the index of the most recently received update
\begin{equation}
N(s)\coloneqq \max\{k|t{'}_{k}\leq s\},
\label{eq:indexN}
\end{equation}
and define the time stamp of the most recently received update
\begin{equation}
U(s)\coloneqq t_{N(s)}.
\end{equation}

Finally, we define the \textit{age of information} as the random process
\begin{equation}
\Delta(s)\coloneqq s-U(s).
\label{eq:ageDef}
\end{equation}
\end{defn}

As mentioned in section \ref{sec:literature}, the characterization of \textit{average age} for status update systems using queuing theory has been considered before, but most previous work assumed the transmission of all generated packets. In this work, we consider that the source node can discard some of the arriving packets, in a process that is referred to as \textit{packet management}. We compare three policies of packet management:

\textbf{(\textit{i})} assume that samples which arrive while a packet is being transmitted are discarded, and the ones that find the source node idle are immediately transmitted to the destination. In this case, no packets are kept in a queue waiting for transmission. Assuming a Poisson arrival process, FCFS policy, and exponentially distributed service time, this scheme is modeled as an M/M/1/1 queue, where the last entry in the Kendall notation refers to the total capacity of the queuing system, which is a single packet in service;

\textbf{(\textit{ii})} assume that a single packet may be kept in queue, waiting for transmission if the server is busy transmitting another packet. If the server is idle, the service starts immediately. Under the assumptions of Poisson arrivals, and exponential service times, this second model of packet management can be studied as an M/M/1/2 queue. Here the total capacity of the queuing system is one packet in service and one packet in queue; 

\textbf{(\textit{iii})} The third packet management policy also assumes that a single packet may be kept in the buffer waiting for transmission. We propose that packets waiting for transmission are replaced upon arrival of a more up-to-date packet. Keeping the other assumptions, we can model this proposed system as a modified M/M/1/2 queue, identified as an M/M/1/2* queue. This is a peculiar queuing model, for which some of the classic results  from queuing theory, such as Little's result \cite[Chapter 2]{Kleinrock1976}, fail to apply.

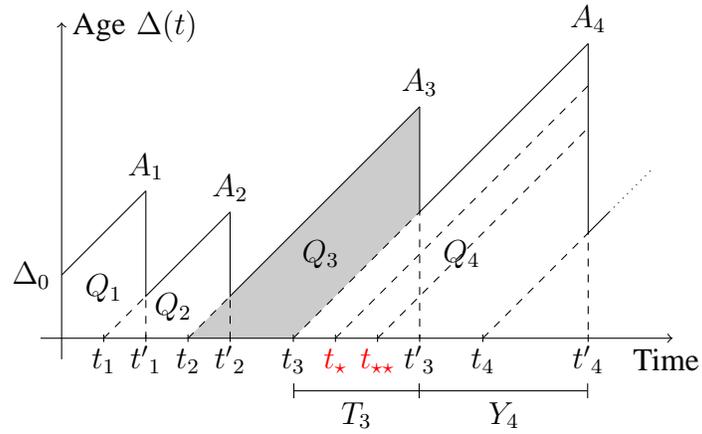
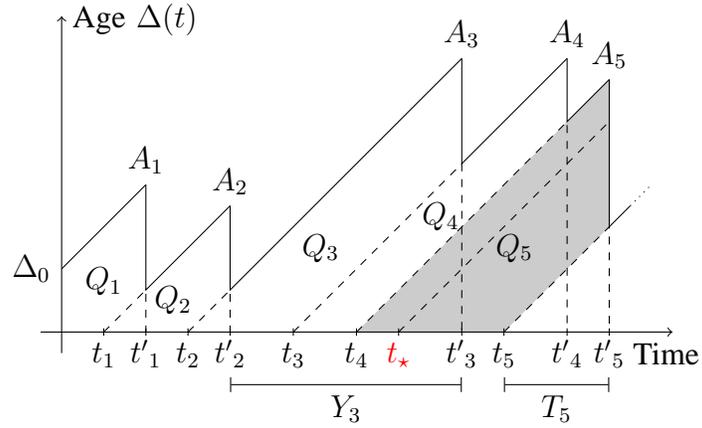
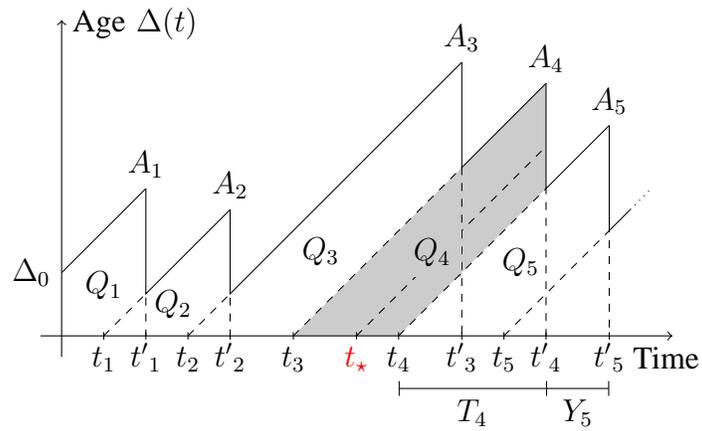
\begin{figure}
    \centering
    \subfigure[Sample path with M/M/1/1 model]
    {
	\begin{tikzpicture}[scale=0.28]
\fill [black!20!white](5,0)--(10,0)--(16,6)--(16,11)--(5,0);
\draw[->](-2,0)--(28,0) node[below, pos=0.99] {Time};
\draw[->](-1,-1)--(-1,15) node[right, pos=0.99] {Age $\Delta(t)$};
\draw(-1,3)--(3,7) node[above]{$A_1$};
\draw(3,7)--(3,2)--(7,6) node[above]{$A_2$};
\draw(7,6)--(7,2)--(16,11) node[above]{$A_3$};
\draw(16,11)--(16,6)--(24,14) node[above]{$A_4$};
\draw(24,14)--(24,5)--(25,6);
\node at (1,2.4) {$Q_1$};
\node at (4.3,1.5) {$Q_2$};
\node at (11.3,4) {$Q_3$};
\node at (18,4) {$Q_4$};
\node [left] at (-1,3) {$\Delta_0$};
\draw[dotted](25,6)--(27,8);
\draw[dashed](1,0)--(3,2)--(3,0);
\draw[dashed](5,0)--(7,2)--(7,0);
\draw[dashed](10,0)--(16,6)--(16,0);
\draw[dashed](12,0)--(16,4)--(24,12);
\draw[dashed](14,0)--(16,2)--(24,10);
\draw[dashed](19,0)--(24,5)--(24,0);
\draw[white] (-1,-2)--(25,-2.2) node [above,color=black] at (1,-2.2){$t_{1}$} node [above,color=black] at (3,-2.2){$t{'}_{1}$} node [above,color=black] at (5,-2.2){$t_{2}$}
node [above,color=black] at (7,-2.2){$t{'}_{2}$} node [above,color=black] at (10,-2.2){$t_{3}$} node [above, color=red] at (12,-2.2){$t_{\star}$} 
node [above, color=red] at (14,-2.2){$t_{\star\star}$} node [above,color=black] at (16,-2.2){$t{'}_{3}$} node [above,color=black] at (19,-2.2){$t_{4}$}
node [above,color=black] at (24,-2.2){$t{'}_{4}$};
\draw (1,5pt) -- (1,-5pt);
\draw (3,5pt) -- (3,-5pt);
\draw (5,5pt) -- (5,-5pt);
\draw (7,5pt) -- (7,-5pt);
\draw (10,5pt) -- (10,-5pt) ;
\draw (12,5pt) -- (12,-5pt);
\draw (14,5pt) -- (14,-5pt);
\draw (16,5pt) -- (16,-5pt);
\draw (19,5pt) -- (19,-5pt);
\draw (24,5pt) -- (24,-5pt);
%\draw(1,-2.5)--(5,-2.5) node[below] at (3,-2.5) {$X_2$};
%\draw[|-](5,-2.5)--(10,-2.5) node[below] at (7.5,-2.5) {$X_3$};
\draw[|-|](10,-2.5)--(16,-2.5) node[below] at (13,-2.5) {$T_3$};
\draw[-|](16,-2.5)--(24,-2.5) node[below] at (20,-2.5) {$Y_4$};
\end{tikzpicture}
	\label{fig:mm11_saw}
    }
    \\
    \subfigure[Sample path with M/M/1/2 model]
    {
	\begin{tikzpicture}[scale=0.28]
\fill [black!20!white](13,0)--(20,0)--(25,5)--(25,12)--(13,0);
\draw[->](-2,0)--(28,0) node[below, pos=0.99] {Time};
\draw[->](-1,-1)--(-1,15) node[right, pos=0.99] {Age $\Delta(t)$};
\draw(-1,3)--(3,7) node[above]{$A_1$};
\draw(3,7)--(3,2)--(7,6) node[above]{$A_2$};
\draw(7,6)--(7,2)--(18,13) node[above]{$A_3$};
\draw(18,13)--(18,8)--(23,13) node[above]{$A_4$};
\draw(23,13)--(23,10)--(25,12) node[above]{$A_5$};
\draw(25,12)--(25,5)--(26,6);
\node at (1,2.4) {$Q_1$};
\node at (4.3,1.5) {$Q_2$};
\node at (11.3,4) {$Q_3$};
\node at (17,5.5) {$Q_4$};
\node at (20.5,4) {$Q_5$};
\node [left] at (-1,3) {$\Delta_0$};
\draw[dashed](1,0)--(3,2)--(3,0);
\draw[dashed](5,0)--(7,2)--(7,0);
\draw[dashed](10,0)--(18,8)--(18,0);
%\draw[dashed](12,0)--(22,10)--(22.0);
\draw[dashed](13,0)--(23,10)--(23,0);
\draw[dashed](15,0)--(25,10);
\draw[dashed](20,0)--(25,5)--(25,0);
\draw[dotted](26,6)--(27,7);
\draw[white] (-1,-2)--(25,-2.2) node [above,color=black] at (1,-2.2){$t_{1}$} node [above,color=black] at (3,-2.2){$t{'}_{1}$} node [above,color=black] at (5,-2.2){$t_{2}$}
node [above,color=black] at (7,-2.2){$t{'}_{2}$} node [above,color=black] at (10,-2.2){$t_{3}$} node [above, color=black] at (13,-2.2){$t_{4}$} 
node [above, color=red] at (15,-2.2){$t_{\star}$} node [above,color=black] at (18,-2.2){$t{'}_{3}$} node [above,color=black] at (20,-2.2){$t_{5}$}
node [above,color=black] at (23,-2.2){$t{'}_{4}$}node [above,color=black] at (25,-2.2){$t{'}_{5}$};
\draw (1,5pt) -- (1,-5pt);
\draw (3,5pt) -- (3,-5pt);
\draw (5,5pt) -- (5,-5pt);
\draw (7,5pt) -- (7,-5pt);
\draw (10,5pt) -- (10,-5pt) ;
\draw (13,5pt) -- (13,-5pt);
\draw (15,5pt) -- (15,-5pt);
\draw (18,5pt) -- (18,-5pt);
\draw (20,5pt) -- (20,-5pt);
\draw (23,5pt) -- (23,-5pt);
\draw (25,5pt) -- (25,-5pt);
%\draw[|-|](1,-2.5)--(5,-2.5) node[below] at (3,-2.5) {$X_2$};
\draw[|-|](7,-2.5)--(18,-2.5) node[below] at (12.5,-2.5) {$Y_3$};
\draw[|-|](20,-2.5)--(25,-2.5) node[below] at (22.5,-2.5) {$T_5$};
\end{tikzpicture}
	\label{fig:mm12_saw}
    }
    \\
    \subfigure[Sample path with M/M/1/2* model]
    {
	\begin{tikzpicture}[scale=0.28]
\fill [black!20!white](10,0)--(15,0)--(22,7)--(22,12)--(10,0);
\draw[->](-2,0)--(28,0) node[below, pos=0.99] {Time};
\draw[->](-1,-1)--(-1,15) node[right, pos=0.99] {Age $\Delta(t)$};
\draw(-1,3)--(3,7) node[above]{$A_1$};
\draw(3,7)--(3,2)--(7,6) node[above]{$A_2$};
\draw(7,6)--(7,2)--(18,13) node[above]{$A_3$};
\draw(18,13)--(18,8)--(22,12) node[above]{$A_4$};
\draw(22,12)--(22,7)--(25,10) node[above]{$A_5$};
\draw(25,10)--(25,5)--(26,6);
\node at (1,2.4) {$Q_1$};
\node at (4.3,1.5) {$Q_2$};
\node at (11.3,4) {$Q_3$};
%\node at (17,5.5) {$Q_4$};
\node at (20.8,3.5) {$Q_5$};
\node [left] at (-1,3) {$\Delta_0$};
\draw[dashed](1,0)--(3,2)--(3,0);
\draw[dashed](5,0)--(7,2)--(7,0);
\draw[dashed](10,0)--(18,8)--(18,0);
\draw[dashed](13,0)--(22,9)  node[pos=0.4, yshift={3pt},fill=black!20!white]{$Q_4$};
\draw[dashed](15,0)--(22,7)--(22,0);
\draw[dashed](20,0)--(25,5)--(25,0);
\draw[dotted](26,6)--(27,7);
\draw[white] (-1,-2)--(25,-2.2) node [above,color=black] at (1,-2.2){$t_{1}$} node [above,color=black] at (3,-2.2){$t{'}_{1}$} node [above,color=black] at (5,-2.2){$t_{2}$}
node [above,color=black] at (7,-2.2){$t{'}_{2}$} node [above,color=black] at (10,-2.2){$t_{3}$} node [above, color=red] at (13,-2.2){$t_{\star}$} 
node [above, color=black] at (15,-2.2){$t_{4}$} node [above,color=black] at (18,-2.2){$t{'}_{3}$} node [above,color=black] at (20,-2.2){$t_{5}$}
node [above,color=black] at (22,-2.2){$t{'}_{4}$}node [above,color=black] at (25,-2.2){$t{'}_{5}$};
\draw (1,5pt) -- (1,-5pt);
\draw (3,5pt) -- (3,-5pt);
\draw (5,5pt) -- (5,-5pt);
\draw (7,5pt) -- (7,-5pt);
\draw (10,5pt) -- (10,-5pt) ;
\draw (13,5pt) -- (13,-5pt);
\draw (15,5pt) -- (15,-5pt);
\draw (18,5pt) -- (18,-5pt);
\draw (20,5pt) -- (20,-5pt);
\draw (22,5pt) -- (22,-5pt);
\draw (25,5pt) -- (25,-5pt);
%\draw[|-|](1,-2.5)--(5,-2.5) node[below] at (3,-2.5) {$X_2$};
\draw[|-](15,-2.5)--(22,-2.5) node[below] at (18.5,-2.5) {$T_4$};
\draw[|-|](22,-2.5)--(25,-2.5) node[below] at (23.5,-2.5) {$Y_5$};
\end{tikzpicture}
	\label{fig:mm12star_saw}
    }
    \caption{Sawtooth curve - Examples of sample path for \textit{age} $\Delta(t)$. We identify examples of the time in the system $T_k$, the interdeparture time $Y_k$, the peaks $A_k$, and the areas of geometric forms $Q_k$, which will be used in the characterization of the \textit{age} throughout this paper.}
    \label{fig:sawtooth3}
\end{figure}

In order to understand more about the process $\Delta(t)$, defined in \eqref{eq:ageDef}, we depict in Figure \ref{fig:sawtooth3} examples of sample paths for each of the models described in \textbf{(\textit{i})}-\textbf{(\textit{iii})}. The \textit{age of information} at the destination node increases linearly with time. Upon reception of a new status update, the \textit{age} is reset to the difference of the current time instant and the time stamp of the received update. Recall that $t_k$ denotes the time instant the $k$th packet was generated, and $t{'}_{k}$ the time instant that this packet completes service. We identify with $T_k$ the time spent in the system, defined as $$T_k\coloneqq t{'}_{k}-t_{k},$$ and let the interdeparture time be denoted with $Y_k$, defined as $$Y_k\coloneqq t{'}_{k}-t{'}_{k-1}.$$

Figure \ref{fig:mm11_saw} illustrates the case \textbf{(\textit{i})} of an M/M/1/1 queue, and the packets arriving at times $t_{\star}$ and $t_{\star\star}$ are discarded, since the server was found busy upon arrival. A sample path example for the case \textbf{(\textit{ii})} is shown in Figure \ref{fig:mm12_saw}, which considers an M/M/1/2 queue model. In this model, there is a single buffer space, and packets that find the server busy and the buffer occupied are discarded, as illustrated for a packet arriving at time $t_{\star}$. Figure \ref{fig:mm12star_saw} presents an example of the evolution of \textit{age} with the policy described in item \textbf{(\textit{iii})}, modeled as an M/M/1/2* queue. No packets are blocked from entering the queue. If a new packet arrives while the system is full, the packet waiting is discarded. In the illustration, the packet that arrived at time $t_{\star}$ spends some time in the buffer, but it is substituted when a new packet arrives at time $t_4$. The transmission of this last packet begins when the server becomes available, at time $t{'}_{3}$. 

Notation: the subindex $k$ will be used to refer to successfully transmitted packets only. We will not identify the time of arrival of discarded packets. As an example, consider the sample path illustrated in Figure \ref{fig:mm11_saw}. The fourth  and fifth packets to arrive are discarded. In this case, the fourth transmitted packet was the sixth packet to arrive, but we use index $k=4$, arrival time $t_4$, and departure time $t{'}_4$. We identify the fourth peak with $A_4$, and the area of the fourth trapezoid with $Q_4$. In general we will write $A_k$, $Q_k$, $Y_k$, $T_k$ to refer to quantities associated with the $k$th transmitted packet. 

\section{Metrics for Age of Information}\label{sec:metrics}

After observing the sawtooth curves in Figure \ref{fig:sawtooth3}, we present in this section two metrics that can be used to characterize the \textit{age of information}, namely the \textit{average age} and the \textit{peak age}.

\subsection{Average Age}
Assuming ergodicity of the process $\Delta(t)$, the \textit{average age} can be calculated using a time average. Consider an observation interval $(0,\tau)$. The \textit{time average age} is 
\begin{equation}
\Delta_{\tau}=\frac{1}{\tau}\int_{0}^{\tau}\Delta(t)dt.
\label{eq:timeAvg}
\end{equation}

The integration in \eqref{eq:timeAvg} can be interpreted as the area under the curve for $\Delta(t)$. Observing Figure \ref{fig:sawtooth3}, note that the area can be calculated as the sum of disjoint geometric parts with areas identified by $Q_k$, $k=1,2,\ldots N(\tau)$, with $N(\tau)$ as defined in \eqref{eq:indexN}. In the case that $\tau > t{'}_{N(\tau)}$, there will also be a partial area to be added in the end, which we will denote with $\tilde{Q}$. Summing all the areas, we write the \textit{time average age} as
\begin{IEEEeqnarray}{rCl}
\Delta_{\tau}&=& \frac{1}{\tau} \left(Q_1+\tilde{Q}+\sum_{k=2}^{N(\tau)}{Q_k}\right)\nonumber\\
&=&\frac{Q_1+\tilde{Q}}{\tau}+\frac{N(\tau)-1}{\tau}\frac{1}{N(\tau)-1}\sum_{k=2}^{N(\tau)}{Q_k}.
\label{eq:Delta_tau}
\end{IEEEeqnarray}

The \textit{average age} is calculated as we take the length of the observation interval to infinity 
\begin{equation}
\Delta = \lim_{\tau\rightarrow\infty}\Delta_{\tau}. 
\label{eq:Delta_avgaslimit}
\end{equation}

The ratio of the number of transmitted packets by the length of the interval converges to the rate of transmitted packets, referred to as the effective arrival rate. We define
\begin{defn}{Effective Arrival Rate}
\begin{equation}
\lambda_e\coloneqq\lim_{\tau\rightarrow\infty}\frac{N(\tau)}{\tau}. 
\label{eq:lambdaeff}
\end{equation}
\end{defn}

Also, as $\tau\rightarrow\infty$, the number of transmitted packets grows to infinity, $N(\tau)\rightarrow\infty$. Due to the ergodicity of $Q_k$, the \textit{average age} can be calculated as
\begin{equation}
\Delta =\lambda_e\mathds{E}[Q_k]
\label{eq:AvgAge}
\end{equation}

To calculate the area, $Q_k$, we take the area of the bigger isosceles triangle with sides $T_{k-1}+Y_k$, and subtract the area of the smaller triangle with sides $T_k$. The average area is 
\begin{IEEEeqnarray}{rCl}
\mathds{E}[Q_k]&=&\frac{1}{2}\mathds{E}[(T_{k-1}+Y_k)^2]-\frac{1}{2}\mathds{E}[T_k^2]\nonumber\\
&=&\frac{1}{2}\mathds{E}[Y_k^2]+\mathds{E}[T_{k-1}Y_k],
\label{eq:EQk}
\end{IEEEeqnarray}
where the second equality follows from the fact that $T_{k-1}$ and $T_k$ are identically distributed. 

\subsection{Peak Age}

Depending on the application, it may be necessary to characterize the maximum value of the \textit{age of information}, immediately before an update is received. It may also be desirable to optimize the system so that the \textit{age} remains below a threshold with a certain probability. 

With that motivation, we propose an alternative metric to the \textit{average age}, obtained observing the peak values in the sawtooth curve. The \textit{peak age} provides information about the worst case \textit{age}, and its expected value can be easily calculated in many cases of interest.

Observe the example sample paths shown in Figure \ref{fig:sawtooth3}. Consider the peak corresponding to the $k$th successfully received packet. The value of \textit{age} in the peak is denoted with $A_k$. We present the definition of \textit{peak age} below.
\begin{defn}{Peak Age}

Let $T_{k-1}$ be the time in the system for the previously transmitted packet, and $Y_k$ be the interdeparture time, or the time elapsed between service completion of the $(k-1)$th packet and service completion of the $k$th packet. The value of \textit{age} achieved immediately before receiving the $k$th update is called \textit{peak age}, and defined as
\begin{equation}
A_k \coloneqq T_{k-1}+Y_k.
\label{eq:avgAk}
\end{equation}
\end{defn}

\section{Characterizing Average Age and Peak Age for Selected Queuing Models}\label{sec:analytic}

In this section we characterize the \textit{average age} and the \textit{peak age} for each of the three queuing models mentioned in \textbf{(\textit{i})}-\textbf{(\textit{iii})}. We start with a discussion about the interdependence between the variables $T_{k-1}$ and $Y_k$.  For the \textit{peak age} we are interested in the distribution of their sum, while for the calculations of the \textit{average age}, we need to obtain the expected values in \eqref{eq:EQk}.

Using queuing system models, we are able to characterize both the time in the system and the interdeparture times but, in general, these two random variables are not independent and we do not have information about the joint distribution. Nonetheless, it is possible to describe an event such that $T_{k-1}$ and $Y_k$ are conditionally independent. 

Let $\psi$ be the event that a packet left behind an empty system upon departure. Under the assumption of Poisson arrivals, the time until the next arrival is exponentially distributed with parameter $\lambda$, due to the memoryless property of interarrival times. The arriving packet will be served immediately, and the service time is assumed to be exponentially distributed with parameter $\mu$. The interarrival and service times are independent, and they are also independent from the time in the system of the packet that just left.

We denote with $\bar{\psi}$ the complement event that the system is not empty upon departure, i.e. there was at least one packet waiting in queue. In this case, the packet waiting in queue starts service immediately, and the time until the next departure is simply a service time, which is independent of the time in the system for the packet previously served.

For queuing systems with single, as opposed to batch, arrivals, the probability of leaving the system empty upon departure is equal to the steady state probability that the system is empty \cite[Chapter 5]{Cooper1981}. The steady state distribution is particular to each queuing model, and will be presented later in this section.

The variables $T_{k-1}$ and $Y_k$ are conditionally independent, given the event $\psi$ that the $(k-1)$th packet leaves behind an empty system upon departure. While the conditional distribution of the time in the system depends on the selected queuing model, the conditional distribution of the interdeparture time can already be stated here. It is given by the convolution of the distributions of two exponential random variables, with parameters $\lambda$ and $\mu$, which yields
\begin{equation}
f(y|\psi)=\frac{\lambda \mu}{\mu-\lambda}\left[e^{-\lambda y}-e^{-\mu y}\right].
\label{eq:YgivenPsi}
\end{equation}
\begin{equation}
\mathds{E}[Y_k|\psi]=\frac{1}{\lambda}+\frac{1}{\mu}, 
\label{eq:YkPsi}
\end{equation}
\begin{equation}
\mathds{E}[Y_k^2|\psi]=\frac{2( \lambda^2 + \lambda \mu + \mu^2)}{\lambda^2 \mu^2}.
\label{eq:Yksq}
\end{equation}

We also have that 
\begin{equation}
f(y|\bar{\psi})=\mu\exp(-\mu y).
\label{eq:YgivenBarPsi}
\end{equation}
\begin{equation}
\mathds{E}[Y_k|\bar{\psi}]=\frac{1}{\mu}, 
\label{eq:YkPsi2}
\end{equation}
\begin{equation}
\mathds{E}[Y_k^2|\bar{\psi}]=\frac{2}{\mu^2}.
\label{eq:Yksq_2}
\end{equation}

The expectation of the product, $\mathds{E}[T_{k-1}Y_{k}]$, to be used in \eqref{eq:EQk}, will be calculated later in this section for each of the queuing models, using nested expectations, as
\begin{IEEEeqnarray}{rCl}
\mathds{E}[T_{k-1}Y_k]&=&\mathds{P}(\psi)\mathds{E}[T_{k-1}Y_k|\psi]+\mathds{P}(\bar{\psi})\mathds{E}[T_{k-1}Y_k|\bar{\psi}]\nonumber\\
&=&\mathds{P}(\psi)\left(\mathds{E}[T_{k-1}|\psi]\mathds{E}[Y_k|\psi]\right)+\mathds{P}(\bar{\psi})\left(\mathds{E}[T_{k-1}|\bar{\psi}]\mathds{E}[Y_k|\bar{\psi}]\right)
\label{eq:ETY}
\end{IEEEeqnarray}

For each queuing model investigated in this paper, we will describe the probability distribution of the \textit{peak age} by first conditioning on the event $\psi$. The conditional probability density function is then obtained as the convolution of the conditional density functions for $T_{k-1}$ and $Y_k$, which we represent as
\begin{equation}
f(a|\psi)=f(t|\psi)\ast f(y|\psi).
\label{eq:PeakGivenPsi}
\end{equation}

Similarly, we obtain the conditional distribution given the event $\bar{\psi}$, and then write the probability distribution of the \textit{peak age} using
\begin{equation}
f(a)=\mathds{P}(\psi)f(a|\psi)+\mathds{P}(\bar{\psi})f(a|\bar{\psi}).
\label{eq:peak_distr}
\end{equation}

In what follows, each subsection describes in detail the calculations of \textit{average age} and \textit{peak age} for the selected queuing models, namely the M/M/1/1, M/M/1/2, and M/M/1/2* models. For each model, we need to describe the steady state distribution of the number of packets in the system, in order to calculate $\mathds{P}(\psi)$. We also need to characterize the distribution of the time in the system, in order to calculate the expectation in \eqref{eq:ETY} and the distribution of the \textit{peak age}.

\subsection{M/M/1/1 Queue}
\subsubsection{Preliminary Calculations}
An M/M/1/1 queue can be described using a two-state Markov chain, with each state representing the server as idle or busy. We denote with $p_0$ the probability of an empty system, and $p_1$ is the probability of one packet in the system. Analyzing the two-state Markov chain, it is straightforward to obtain \cite[Chapter 3]{Kleinrock1976}
\begin{equation}\label{eq:ststate_mm11}
p_0=\frac{\mu}{\lambda+\mu}; \;\;p_1=\frac{\lambda}{\lambda+\mu}.
\end{equation}   

In the case of an M/M/1/1 queue, a packet is accepted in the system only if the server is idle. The effective arrival rate will be denoted with $\lambda_e$, and we have $\lambda_e=\lambda (1-p_1)$. The time in the system for transmitted packets is equal to the service time, which is assumed to be exponentially distributed with mean $1/\mu$. The average number of packets in the system is $\mathds{E}[N]=0p_0+1p_1$. 

\subsubsection{Average Age}
Recall that $\psi$ represents the event that a transmitted packet left behind an empty system upon departure. In the M/M/1/1 model, this is a certain event, since there is only one packet at a time in the system. In this model, the time in the system for transmitted packets is independent of the interdeparture time. Using this fact in \eqref{eq:EQk}, together with \eqref{eq:YkPsi} and \eqref{eq:Yksq}, the \textit{average age} for the M/M/1/1 model is calculated as
\begin{IEEEeqnarray}{rCl}
\Delta_{M/M/1/1}&=&\lambda_e\mathds{E}[Q_k]\nonumber\\
&=&\lambda_e\left(\frac{1}{2}\mathds{E}[(Y_k)^2]+\mathds{E}[T_{k-1}]\mathds{E}[Y_k]\right)\nonumber\\
&=&\frac{\lambda \mu}{\lambda+\mu}\left[\left(\frac{\lambda+\mu}{\lambda \mu}\right)^2-\frac{1}{\lambda \mu}+\frac{1}{\mu}\frac{\lambda+\mu}{\lambda \mu}\right]\nonumber\\
&=&\frac{1}{\lambda}+\frac{2}{\mu}-\frac{1}{\lambda+\mu}.
\label{eq:Age_mm11}
\end{IEEEeqnarray}

The limit as the arrival rate goes to infinity is
\begin{equation}
\lim_{\lambda \rightarrow\infty}\Delta_{M/M/1/1}=\frac{2}{\mu},
\end{equation}
which is equal to the lower bound for the average age among all first-come-first-served  (FCFS) systems with a single server \cite{Kaul2012a}. In other words, regarding the \textit{average age}, this scheme is asymptotically optimal in this class of models.

\subsubsection{Peak Age}
For the \textit{peak age}, the probability density function is given by the convolution of an exponential distribution for the service time with the distribution of the interdeparture times shown in \eqref{eq:YgivenPsi}. As a result, we have
\begin{IEEEeqnarray}{rCl}
f(a)_{M/M/1/1}&=&f(a|\psi)\nonumber\\
&=&f(t|\psi)\ast f(y|\psi)\nonumber\\
&=&\left( \frac{\mu}{\lambda-\mu}\right)^{2}
\left(\lambda e^{-\lambda a}-\lambda e^{-\mu a}+\lambda(\lambda-\mu)ae^{-\mu a}\right)
\label{eq:PeakMM11}
\end{IEEEeqnarray}

The complementary cumulative distribution function, which describes the probability that the \textit{peak age} surpasses a threshold, is described as
\begin{equation}
\mathds{P}(A>a)_{M/M/1/1}=\left(\frac{\mu}{\lambda-\mu}\right)^2 e^{-\lambda a}+\left[1-\left(\frac{\mu}{\lambda-\mu}\right)^2\right]e^{-\mu a}+\frac{\lambda \mu}{\lambda-\mu}a e^{-\mu a}
\label{eq:PeakComplementaryMM11}
\end{equation}

The average \textit{peak age} for the M/M/1/1 model can be obtained integrating \eqref{eq:PeakComplementaryMM11}. It can also be calculated easily, observing that $\mathds{E}[T_{k-1}]=1/\mu$, while $\mathds{E}[Y_k]=\mathds{E}[Y_k|\psi]$, as in \eqref{eq:YkPsi}. As a result, 
\begin{IEEEeqnarray}{rCl}
\mathds{E}[A_k]_{M/M/1/1} &=&\frac{1}{\mu}+\frac{1}{\lambda}+\frac{1}{\mu}\nonumber\\
&=&\frac{1}{\lambda}+\frac{2}{\mu}.
\label{eq:avgAk-mm11}
\end{IEEEeqnarray}

\subsection{M/M/1/2 Queue}
\subsubsection{Preliminary Calculations}
The M/M/1/2 queue can be described by a Markov chain with three states, that represent an empty system, a single packet being served, or a packet in service with a packet waiting in the buffer. Let $\rho\coloneqq\lambda/\mu$. The analysis of the three-state Markov chain yields the steady state probabilities \cite[Chapter 3]{Kleinrock1976}
\begin{equation}
p_j=\frac{\rho^j}{1+\rho+\rho^2}, \;\;j\in\{0,1,2\}.
\label{eq:ss_prob}
\end{equation}

Once we have the steady state distribution, we can calculate $\mathds{P}(\psi)$, which is equal to the steady state probability that the system is empty \cite[Chapter 5]{Cooper1981}. We normalize the probabilities and write
\begin{IEEEeqnarray}{rClCl}
\mathds{P}(\psi)&=&\frac{p_0}{p_0+p_1}&=&\frac{\mu}{\lambda+\mu}, \label{eq:P_phi}\\
\mathds{P}(\bar{\psi})&=&\frac{p_1}{p_0+p_1} &=&\frac{\lambda}{\lambda+\mu}.\label{eq:P_phi_2}
\end{IEEEeqnarray}

A packet is accepted in the system as long as the system is not full. Therefore, the effective arrival rate is $\lambda_e=\lambda(1-p_2)$. 

We can already calculate the first term of $\mathds{E}[Q_k]$ using \eqref{eq:Yksq}, \eqref{eq:Yksq_2}, together with \eqref{eq:P_phi} and \eqref{eq:P_phi_2}, and we have
\begin{IEEEeqnarray}{rCl}
\frac{1}{2}\mathds{E}[Y_k^2]&=&\frac{1}{2}\left(\mathds{E}[Y_k^2|\psi]\mathds{P}(\psi)+\mathds{E}[Y_k^2|\bar{\psi}]\mathds{P}(\bar{\psi}) \right)\nonumber\\
&=&\frac{1}{2}\left(\frac{2(\lambda^2+\lambda \mu+\mu^2)}{(\lambda \mu)^2}\frac{\mu}{\lambda+\mu}+\frac{2}{\mu^2}\frac{\lambda}{\lambda+\mu}\right) \nonumber\\
&=&\frac{1}{\lambda^2}+\frac{1}{\mu^2}.
\label{eq:EY2}
\end{IEEEeqnarray}

Next we need to characterize the time in the system for the transmitted packets, and also its conditional distribution, given the event $\psi$. The time in the system can be written as $T_{k-1}=W_{k-1}+S_{k-1}$, where $W_{k-1}$ is a random variable representing the waiting time, and $S_{k-1}$ represents the service time.

The waiting time is zero if the system is found idle upon arrival. Due to the PASTA property (Poisson Arrivals See Time Averages) \cite[Chapter 3]{Cooper1981}, the probability that the system is found idle upon arrival is equal to the steady state probability that the system is idle. If the system is found busy upon arrival, the waiting time is the remaining service time, which is exponentially distributed with parameter $\mu$, due to the memoryless property of service times. Note that the waiting time is independent of future arrivals, hence independent of the event $\psi$. For the waiting time of a transmitted packet, we have
\begin{equation}
\mathds{P}(W_{k-1}>w)=\frac{p_1}{p_0+p_1}e^{-\mu w}, \; w>0.
\label{eq:WGw}
\end{equation}

The average waiting time can be obtained integrating \eqref{eq:WGw}, or using Little's result \cite[Chapter 2]{Kleinrock1976} applied to the average number of packets in the buffer, $\mathds{E}[N_q]=1p_2$, and using the effective arrival rate, which yields
\begin{IEEEeqnarray}{rCl}
\mathds{E}[W_{k-1}]&=&\frac{\mathds{E}[N_q]}{\lambda_e}\nonumber\\
&=&\frac{1 p_2}{\lambda(1-p_2)}\nonumber\\
&=&\frac{\lambda}{\mu(\lambda+\mu)}
\label{eq:Wtx_mm12}
\end{IEEEeqnarray}

For the time in the system, we can condition on the state found upon arrival (idle or busy), separating the cases with and without waiting. For transmitted packets we have
\begin{IEEEeqnarray}{rCl}
\mathds{P}(T_{k-1}>t)&=&\frac{p_0}{p_0+p_1}\mathds{P}(S_{k-1}>t|\text{idle})+\frac{p_1}{p_0+p_1}\mathds{P}(S_{k-1}+W_{k-1}>t|\text{busy})\nonumber\\
&=&\frac{\mu}{\lambda+\mu} e^{-\mu t}+\frac{\lambda}{\lambda+\mu}e^{-\mu t}\left(1+ \mu t \right),
\label{eq:TGt}
\end{IEEEeqnarray}
where the second equality follows from the fact that, given that the server was found busy upon arrival, $W_{k-1}$ and $S_{k-1}$ are two independent random variables, each exponentially distributed with parameter $\mu$.   

The average time in the system can be calculated integrating the result in \eqref{eq:TGt}. It is also straightforward to obtain the average time in the system using \eqref{eq:Wtx_mm12} and adding the average service time. Alternatively, it can be calculated using Little's result with the average total number of packets in the system, $\mathds{E}[N]=0p_0+1p_1+2p_2$, and the effective arrival rate, which yields
\begin{equation}
\mathds{E}[T_{k-1}]=\frac{\mu+2\lambda}{\mu(\lambda+\mu)}.
\label{eq:ETTx_mm12}
\end{equation}

We also need the conditional distribution $f(t|\psi)$ of the time in the system given the event $\psi$. We note that the $(k-1)$th transmitted packet leaves the system idle upon departure if and only if zero arrivals occur while the $(k-1)$th transmitted packet is being served. Let $f(s)$ be the probability density function of the service time. We define the probability $\mathds{P}(\psi|S_{k-1}=s)$ by requiring that for every measurable set $A\subset [0,\infty)$, we have
\begin{equation}
\mathds{P}(\psi,S_{k-1}\in A)=\int_{A}f(s)\mathds{P}(\psi|S_{k-1}=s)ds.
\end{equation}
The conditional distribution of service time, given that no arrivals occur is calculated as
\begin{IEEEeqnarray}{rCl}
f(s|\psi)&=&\frac{\mathds{P}(\psi|S_{k-1}=s)f(s)}{\int_{0}^{\infty}\mathds{P}(\psi|S_{k-1}=s)f(s)ds}\nonumber\\
&=&\frac{\frac{(\lambda s)^0}{0!}e^{-\lambda s} \mu e^{-\mu s}}{\int_{0}^{\infty}\frac{(\lambda s)^0}{0!}e^{-\lambda s} \mu e^{-\mu s}ds}\nonumber\\
&=&(\lambda+\mu)e^{-(\lambda+\mu)s},
\label{eq:conditionalS}
\end{IEEEeqnarray}
with the conditional expectation
\begin{equation}
\mathds{E}[S_{k-1}|\psi]=\frac{1}{\lambda+\mu}.
\label{eq:ES_psi}
\end{equation}

For the conditional distribution of the time in the system, we use the results in \eqref{eq:WGw} and \eqref{eq:conditionalS} to write
\begin{IEEEeqnarray}{rCl}
\mathds{P}(T_{k-1}>t|\psi)&=&\mathds{P}(W_{k-1}+S_{k-1}>t|\psi)\\
&=&\int_{0}^{\infty}\mathds{P}(W_{k-1}>t-s)f(s|\psi)ds\\
&=&e^{-\mu t},
\label{eq:TGpsi}
\end{IEEEeqnarray}
with conditional expectation
\begin{equation}
\mathds{E}[T_{k-1}|\psi]=\frac{1}{\mu}.
\label{eq:ET_psi}
\end{equation}

We follow similar steps conditioning on the complement event $\bar{\psi}$. The system is left behind with another packet waiting for transmission if and only if at least one arrival occurs during the service time of the $(k-1)$th transmitted packet. The conditional distribution of the service time given the event $\bar{\psi}$ is calculated similarly to \eqref{eq:conditionalS} and yields
\begin{IEEEeqnarray}{rCl}
f(s|\bar{\psi})&=&\frac{\mathds{P}(\bar{\psi}|S_{k-1}=s)f(s)}{\int_{0}^{\infty}\mathds{P}(\bar{\psi}|S_{k-1}=s)f(s)ds}\nonumber\\
&=&\frac{\left(1-\frac{(\lambda s)^0}{0!}e^{-\lambda s}\right) \mu e^{-\mu s}}{\int_{0}^{\infty}\left(1-\frac{(\lambda s)^0}{0!}e^{-\lambda s} \right)\mu e^{-\mu s}ds}\nonumber\\
&=&\frac{\lambda+\mu}{\lambda}\left(1-e^{-\lambda s}\right)\mu e^{-\mu s},
\label{eq:conditionalS_2}
\end{IEEEeqnarray}
with the resulting conditional expectation
\begin{equation}
\mathds{E}[S_{k-1}|\bar{\psi}]=\frac{1}{\mu}+\frac{1}{(\lambda+\mu)}.
\label{eq:ES_psi_2}
\end{equation}

For the time in the system, we write
\begin{IEEEeqnarray}{rCl}
\mathds{P}(T_{k-1}>t|\bar{\psi})&=&\mathds{P}(W_{k-1}+S_{k-1}>t|\bar{\psi})\nonumber\\
&=&\int_{0}^{\infty}\mathds{P}(W_{k-1}>t-s)f(s|\bar{\psi})ds\nonumber\\
&=&e^{-\mu t}(1+\mu t),
\label{eq:TGbarpsi}
\end{IEEEeqnarray}
indicating that the probability distribution in this case is the result of a convolution of two exponential distributions with parameter $\mu$. The conditional expectation is
\begin{equation}
\mathds{E}[T_{k-1}|\bar{\psi}]=\frac{2}{\mu}.
\label{eq:ET_psi_2}
\end{equation}

\subsubsection{Average Age}

Using the probabilities for the events $\psi$ and $\bar{\psi}$ as described in \eqref{eq:P_phi}, and \eqref{eq:P_phi_2}, together with the conditional expectations calculated in \eqref{eq:YkPsi}, and \eqref{eq:YkPsi2} for $Y_k$, and in \eqref{eq:ET_psi} and \eqref{eq:ET_psi_2} for $T_{k-1}$ we can finally calculate the expected value of the product $T_{k-1}Y_k$, to be used in the calculations of the \textit{average age}. We have
\begin{IEEEeqnarray}{rCl}
\mathds{E}[T_{k-1}Y_k]&=&\mathds{E}[T_{k-1}Y_k|\psi]\mathds{P}(\psi)+\mathds{E}[T_{k-1}Y_k|\bar{\psi}]\mathds{P}(\bar{\psi})\nonumber\\
&=&\left(\frac{1}{\lambda}+\frac{1}{\mu}\right)\mathds{E}[T_{k-1}|\psi]\frac{\mu}{\lambda+\mu}+\frac{1}{\mu}\mathds{E}[T_{k-1}|\bar{\psi}]\frac{\lambda}{\lambda+\mu}\nonumber\\
&=&\left(\frac{1}{\lambda}+\frac{1}{\mu}\right)\frac{1}{\mu}\frac{\mu}{\lambda+\mu}+\frac{1}{\mu}\frac{2}{\mu}\frac{\lambda}{\lambda+\mu}\nonumber\\
&=&\frac{2\lambda^2+\lambda \mu +\mu^2}{\lambda\mu^2(\lambda+\mu)}.
\label{eq:ETY_2}
\end{IEEEeqnarray}

We write the \textit{average age} for the M/M/1/2 model using \eqref{eq:EY2} and \eqref{eq:ETY_2} as
\begin{IEEEeqnarray}{rCl}
\Delta_{M/M/1/2}&=&\lambda_e\mathds{E}[Q_k]\nonumber\\
&=&\lambda_e\left(\frac{1}{2}\mathds{E}[(Y_k)^2]+\mathds{E}[T_{k-1}Y_k]\right)\nonumber\\
&=&\frac{\lambda \mu(\lambda+\mu)}{\lambda^2+\lambda \mu +\mu^2}\left(\frac{1}{\lambda^ 2}+\frac{1}{\mu^ 2}+\frac{2\lambda^2+\lambda \mu +\mu^2}{\lambda\mu^2(\lambda+\mu)}\right)\nonumber\\
&=&\frac{1}{\lambda}+\frac{3}{\mu}-\frac{2(\lambda+\mu)}{\lambda^2+\lambda \mu+\mu^2}
\label{eq:Age_mm12}
\end{IEEEeqnarray}

The limit as the arrival rate goes to infinity is
\begin{equation}
\lim_{\lambda \rightarrow\infty}\Delta_{M/M/1/2}=\frac{3}{\mu},
\end{equation}
which is much larger than the case modeled as M/M/1/1. Intuitively, for very large arrival rates, keeping a packet in the buffer is not advantageous, because the information ages while waiting in queue, and it is better to wait for a new packet to arrive instead of keeping one in the buffer.

\subsubsection{Peak Age}
The probability distribution of the \textit{peak age} is calculated conditioning on the events $\psi$ and $\bar{\psi}$. In this case, $T_{k-1}$ and $Y_k$ are conditionally independent random variables, and the distribution of their sum can be obtained using the convolution of their individual conditional probability density functions. Using the probability distributions described in \eqref{eq:YgivenPsi}, \eqref{eq:YgivenBarPsi}, \eqref{eq:TGpsi}, and \eqref{eq:TGbarpsi}, we finally obtain
\begin{IEEEeqnarray}{rCl}
f(a|\psi)_{M/M/1/2}&=&\left(\frac{\mu}{\lambda-\mu}\right)^2\left(\lambda e^{-\lambda a}-\lambda e^{-\mu a}+\lambda(\lambda-\mu)a e^{-\mu a}\right)\label{eq:AGpsi_mm12}\\
f(a|\bar{\psi})_{M/M/1/2}&=&\frac{1}{2}a^2\mu^ 3 e^{-\mu a}\label{eq:AGbarpsi_mm12},
\end{IEEEeqnarray}
and the distribution of the \textit{peak age} is the mixture, as described in \eqref{eq:peak_distr}, using the probabilities in \eqref{eq:P_phi} and \eqref{eq:P_phi_2}.

The complementary cumulative distribution function which describes the probability that the \textit{peak age} surpasses a threshold is
\begin{IEEEeqnarray}{rcl}
&&\mathds{P}(A>a)_{M/M/1/2}=\frac{\mu^3}{(\lambda-\mu)^2(\lambda+\mu)}e^{-\lambda a}\nonumber\\
&&+\frac{\lambda}{2(\lambda-\mu)^2(\lambda+\mu)}e^{-\mu a}\left[\mu^2 a^2 (\lambda-\mu)^2+2 \lambda \mu (\lambda-\mu) a+2(\lambda^2-\lambda\mu-\mu^2) \right]
\label{eq:PeakComplementaryMM12}
\end{IEEEeqnarray}

The expected value of the \textit{peak age} can be calculated directly as
\begin{IEEEeqnarray}{rCl}
\mathds{E}[T_{k-1}]&=&\frac{\mu+2 \lambda}{\mu(\lambda+\mu)}\nonumber\\
\mathds{E}[Y_k]&=&\frac{1}{\lambda}+\frac{\lambda}{\mu(\lambda+\mu)}\nonumber\\
\mathds{E}[A_k]_{M/M/1/2}&=&\mathds{E}[T_{k-1}]+\mathds{E}[Y_k]\nonumber\\
&=&\frac{1}{\lambda}+\frac{3}{\mu}-\frac{2}{\lambda+\mu}
\end{IEEEeqnarray}

\subsection{M/M/1/2* Queue}
\subsubsection{Preliminary Calculations}

The M/M/1/2* is a peculiar queue model, in which the packet waiting in queue is replaced if a new packet arrives. Regarding the number of packets in the system, this model behaves exactly as the M/M/1/2 queue. That is because a replacement only occurs when an arriving packet finds the system full, so the replacement results in a packet being discarded. Intuitively, discarding the packet that just arrived or the packet that was already in the buffer does not change the number of packets in the system. As a result, the M/M/1/2* queue can be described with a three-state Markov chain, and the steady state probabilities in \eqref{eq:ss_prob} still hold. The average total number of packets in the system is $\mathds{E}[N]=0p_0+1p_1+2p_2$, and the effective arrival rate is  $\lambda_e=(1-p_2)\lambda$, as in the M/M/1/2 model.

The first term in the calculation of $\mathds{E}[Q_k]$ , which is $\mathds{E}[Y_k^ 2]/2$ also remains the same as calculated in \eqref{eq:EY2} for the M/M/1/2 model.

It remains to characterize the time in the system for transmitted packets. In the M/M/1/2* model, some packets leave the system after spending some time waiting in the buffer, while other packets end up being transmitted. As a result, Little's result fails to apply to the packets in the buffer. In order to characterize the time in the system for transmitted packets, let $T$ (without any index) be the time in the system for any packet, which may or may not be discarded. Let $\mathds{P}(T>t)$ describe the probability that a packet stays in the system for a time longer than $t$. Conditioning on the state of the server upon arrival (idle or busy), and on the event that the packet is transmitted ($\text{tx}$) or dropped ($\text{drop}$), we calculate 
\begin{IEEEeqnarray}{rCl}
\mathds{P}(T>t)&=&\mathds{P}(\text{idle})\mathds{P}(T>t|\text{idle})\nonumber\\
&&+\mathds{P}(\text{busy, tx})\mathds{P}(T>t|\text{busy, tx})\nonumber\\
&&+\mathds{P}(\text{busy, drop})\mathds{P}(T>t|\text{busy, drop}).
\label{eq:T_1}
\end{IEEEeqnarray}
In what follows, we present the main arguments to calculate each term in \eqref{eq:T_1}.

An arriving packet finds the server idle with probability $\mathds{P}(\text{idle})=p_0$, due to the PASTA property (Poisson Arrivals See Time Averages) \cite[Chapter 3]{Cooper1981}. In this case, the packet is served immediately, and $\mathds{P}(T>t|\text{idle})=e^{-\mu t}$. 

A packet finds the server busy with probability $\mathds{P}(\text{busy})=1-p_0$. According to the packet management scheme, it will be admitted in the system, but it will be transmitted if and only if no other arrival occurs while the service in progress is not completed. Let $R$ represent a remaining service time, with probability density function  $f(r)$, and let $\phi$ be the event that no other packet arrives during the remaining service. We define the probability $\mathds{P}(\phi|R=r)$ by requiring that for every measurable set $A\subset [0,\infty)$, we have
\begin{equation*}
\mathds{P}(\phi,R\in A)=\int_{A}f(r)\mathds{P}(\phi|R=r)dr.
\end{equation*}

The probability of transmission, given that the server was busy upon arrival is calculated as 
\begin{IEEEeqnarray}{rCl}
\mathds{P}(\text{tx}|\text{busy})&=&\int_{0}^{\infty}\mathds{P}(\phi|R=r)f(r)dr\nonumber\\
&=&\int_{0}^{\infty}\frac{(\lambda r)^0}{0!}e^{-\lambda r} \mu e^{-\mu r} dr\nonumber\\
&=&\frac{\mu}{\lambda+\mu}.
\label{eq:TxgivenBusy}
\end{IEEEeqnarray}

As a result, 
\begin{IEEEeqnarray}{rCl}
\mathds{P}(\text{busy,tx})&=&(1-p_0)\frac{\mu}{\lambda+\mu}\\
\mathds{P}(\text{busy,drop})&=&(1-p_0)\frac{\lambda}{\lambda+\mu}
\end{IEEEeqnarray}

The time in the system for a transmitted packet can be written as the sum of waiting and service times, $T=W+S$. Conditioned on the event $\{\text{busy, tx}\}$, the waiting time is distributed as the conditional distribution of the remaining service, given the event $\phi$, calculated as
\begin{IEEEeqnarray}{rCl}
f(w|\text{busy, tx})&=&f(r|\phi)\nonumber\\
&=&\frac{\mathds{P}(\phi|R=r)f(r)}{\int_{0}^{\infty}\mathds{P}(\phi|R=r)f(r)dr}\nonumber\\
&=&\frac{\frac{(\lambda r)^0}{0!} e^{-\lambda r} \mu e^{-\mu r}}{\int_{0}^{\infty}\frac{(\lambda r)^0}{0!}e^{-\lambda r} \mu e^{-\mu r}dr}\nonumber\\
&=&(\lambda+\mu) e^{-(\lambda+\mu)r}.
\label{eq:conditionalWait}
\end{IEEEeqnarray}

Hence, conditioned on the event $\{\text{busy, tx}\}$, $W$ is exponentially distributed with parameter $(\lambda+\mu)$. The conditional probability density function of the time in the system is given by the convolution of the individual densities, since $W$ and $S$ are independent. The corresponding complementary cumulative distribution function we are looking for is
\begin{equation}
\mathds{P}(T>t|\text{busy, tx})=\frac{\lambda+\mu}{\lambda} e^{-\mu t}-\frac{\mu}{\lambda} e^{-(\lambda+\mu)t}
\label{eq:T_busyTx}
\end{equation}

If the packet finds the server busy and is dropped while waiting, the time it spends in the system is exactly the time until the next arrival, to be denoted with $X$, which is exponentially distributed with parameter $\lambda$, conditioned on the event that the next arrival occurs before the end of the service in progress $R$. As a result, the conditional distribution can be calculated as
\begin{IEEEeqnarray}{rCl}
\mathds{P}(T>t|\text{busy, drop})&=&\mathds{P}(X>t|X<R)\nonumber\\
&=&\frac{\mathds{P}(X>t,X<R)}{\mathds{P}(X<R)}\nonumber\\
&=& \frac{\int_{t}^{\infty}(e^{-\lambda t}- e^{-\lambda r})\mu e^{-\mu r}dr}{\int_{0}^{\infty}(1-e^{-\lambda r})\mu e^{-\mu r}dr}\nonumber\\
&=&e^{-(\lambda+\mu)t}
\label{eq:Tdrop}
\end{IEEEeqnarray}

Finally, using the arguments above, we rewrite \eqref{eq:T_1} as
\begin{IEEEeqnarray}{rCl}
\mathds{P}(T>t)&=&p_0\exp(-\mu t)\nonumber\\
&&+(1-p_0)\frac{\mu}{\lambda+\mu}\left[\frac{\lambda+\mu}{\lambda}e^{-\mu t}-\frac{\mu}{\lambda}e^{-(\lambda+\mu)t}\right]\nonumber\\
&&+(1-p_0)\frac{\lambda}{\lambda+\mu}e^{-(\lambda+\mu)t}.
\label{eq:T_2}
\end{IEEEeqnarray}

From \eqref{eq:T_2}, we also have 
\begin{IEEEeqnarray}{rCl}
\mathds{P}(T>t|\text{tx})&=&\frac{p_0}{p_0+(1-p_0)\frac{\mu}{\lambda+\mu}}e^{-\mu t}\nonumber\\
&&+\frac{(1-p_0)\frac{\mu}{\lambda+\mu}}{p_0+(1-p_0)\frac{\mu}{\lambda+\mu}}\left[\frac{\lambda+\mu}{\lambda}e^{-\mu t}-\frac{\mu}{\lambda}e^{-(\lambda+\mu)t}\right],
\label{eq:Ttx}
\end{IEEEeqnarray}
and the expected value
\begin{equation}
\mathds{E}[T|\text{tx}]=\frac{1}{\mu}+\frac{\lambda}{(\lambda+\mu)^2}.
\label{eq:ETTx_mm12drop}
\end{equation}

The expected value of the waiting time for a transmitted packet is calculated using \eqref{eq:conditionalWait} to be
\begin{IEEEeqnarray}{rCl}
\mathds{E}[W|\text{tx}]&=&\frac{(1-p_0)\frac{\mu}{\lambda+\mu}}{p_0+(1-p_0)\frac{\mu}{\lambda+\mu}}\left(\frac{1}{\lambda+\mu}\right)\nonumber\\
&=&\frac{\lambda}{(\lambda+\mu)^2}.
\label{eq:Wtx_drop}
\end{IEEEeqnarray}

\subsubsection{Average Age}

The \textit{average age} for the M/M/1/2* queue is calculated similarly to the M/M/1/2 model. We write the conditional expectations for $T_{k-1}$ using the equality $T_{k-1}=W_{k-1}+S_{k-1}$, with \eqref{eq:ES_psi} and \eqref{eq:ES_psi_2} for the expected service time, with expected waiting time for a transmitted packet is given in \eqref{eq:Wtx_drop}. The resulting conditional expectations for $T_{k-1}$ are
\begin{IEEEeqnarray}{rCl}
\mathds{E}[T_{k-1}|\psi]&=&\mathds{E}[W_{k-1}|\psi]+\mathds{E}[S_{k-1}|\psi]\nonumber\\
&=&\mathds{E}[W_{k-1}]+\mathds{E}[S_{k-1}|\psi]\nonumber\\
&=&\frac{\lambda}{(\lambda+\mu)^2}+\frac{1}{\lambda+\mu}\nonumber\\
&=&\frac{2\lambda+\mu}{(\lambda+\mu)^2},
\label{eq:ET_phi_3}
\end{IEEEeqnarray}
\begin{IEEEeqnarray}{rCl}
\mathds{E}[T_{k-1}|\bar{\psi}]&=&\mathds{E}[W_{k-1}|\bar{\psi}]+\mathds{E}[S_{k-1}|\bar{\psi}]\nonumber\\
&=&\mathds{E}[W_{k-1}]+\mathds{E}[S_{k-1}|\bar{\phi}]\nonumber\\
&=&\frac{\lambda}{(\lambda+\mu)^2}+\frac{1}{\mu}+\frac{1}{(\lambda+\mu)}\nonumber\\
&=&\frac{1}{\mu}+\frac{2\lambda+\mu}{(\lambda+\mu)^2}.
\label{eq:ET_phi_4}
\end{IEEEeqnarray}

Using \eqref{eq:ET_phi_3} and \eqref{eq:ET_phi_4} , together with  the probabilities for the events $\psi$ and $\bar{\psi}$ as described in \eqref{eq:P_phi}, and \eqref{eq:P_phi_2}, and with the conditional expectations calculated in \eqref{eq:YkPsi}, and \eqref{eq:YkPsi2} for $Y_k$, we obtain the expectation of the product $T_{k-1}Y_k$, in the case of an M/M/1/2* model,
\begin{IEEEeqnarray}{rCl}
\mathds{E}[T_{k-1}Y_k]&=&\mathds{E}[T_{k-1}Y_k|\psi]\mathds{P}(\psi)+\mathds{E}[T_{k-1}Y_k|\bar{\psi}]\mathds{P}(\bar{\psi})\nonumber\\
&=&\left(\frac{1}{\lambda}+\frac{1}{\mu}\right)\frac{2\lambda+\mu}{(\lambda+\mu)^2}\frac{\mu}{\lambda+\mu}+\frac{1}{\mu}\left(\frac{1}{\mu}+\frac{2\lambda+\mu}{(\lambda+\mu)^2}\right)\frac{\lambda}{\lambda+\mu}\nonumber\\
&=&\frac{1}{\mu^2}+\frac{1}{\lambda \mu}-\frac{2\lambda+\mu}{(\lambda+\mu)^3}.
\label{eq:ETY_drop}
\end{IEEEeqnarray}

We write the \textit{average age} for the M/M/1/2* model using \eqref{eq:EY2} and \eqref{eq:ETY_drop} as
\begin{IEEEeqnarray}{rCl}
\Delta_{M/M/1/2^{*}}&=&\lambda_e\mathds{E}[Q_k]\nonumber\\
&=&\lambda_e\left(\frac{1}{2}\mathds{E}[(Y_k)^2]+\mathds{E}[T_{k-1}Y_k]\right)\nonumber\\
&=&\frac{\lambda \mu(\lambda+\mu)}{\lambda^2+\lambda \mu +\mu^2}\left(\frac{1}{\lambda^ 2}+\frac{1}{\mu^ 2}+\frac{1}{\mu^2}-\frac{1}{\lambda \mu}-\frac{2\lambda+\mu}{(\lambda+\mu)^3}\right)\nonumber\\
&=& \frac{1}{\lambda} + \frac{2}{\mu} + \frac{\lambda}{(\lambda + \mu)^2} + \frac{1}{\lambda + \mu} - \frac{2 (\lambda+ \mu)}{\lambda^2 + \lambda \mu + \mu^2}
\label{eq:Age_mm12_drop}
\end{IEEEeqnarray}

The limit as the arrival rate goes to infinity is
\begin{equation}
\lim_{\lambda \rightarrow\infty}\Delta_{M/M/1/2^{*}}=\frac{2}{\mu},
\end{equation}
indicating that the model with packet replacement behaves as the model without buffer (M/M/1/1) in the limit. Intuitively, as the arrival rate goes to infinity, a fresh packet will always be available for transmission in both models. The M/M/1/2* model is asymptotically optimal, as the M/M/1/1, but for finite arrival rates and a fixed service rate, the packet replacement provides better performance with respect to the average age, as will be illustrated in section \ref{sec:numerical}.

\subsubsection{Peak Age}

We present the conditional distribution of the \textit{peak age}, given the events $\psi$ and $\bar{\psi}$. The probability density function is obtained as the convolution of the conditional distributions for $T_{k-1}$ and $Y_{k}$, since these variables are conditionally independent. The distribution of the \textit{peak age} is the mixture, as described in \eqref{eq:peak_distr}, using the probabilities in \eqref{eq:P_phi} and \eqref{eq:P_phi_2}.

\begin{IEEEeqnarray}{rCl}
f(a|\psi)_{M/M/1/2^{*}}&=&\left(\lambda (\lambda + \mu) a -\frac{ (2 \mu^3 - \lambda^3 - \lambda^2 \mu)}{\mu (\lambda - \mu)} \right)e^{- (\lambda +\mu)a} \nonumber\\
&&+ \frac{ \lambda \mu + 2 \mu^2}{\lambda- \mu} e^{-\mu a} - \frac{\lambda \mu (\lambda + \mu) + \lambda^3}{\mu (\lambda - \mu)} e^{-\lambda a}
\label{eq:fa_psi}
\end{IEEEeqnarray}
\begin{IEEEeqnarray}{rCl}
f(a|\bar{\psi})_{M/M/1/2^{*}}&=&\frac{\mu^2}{\lambda^2}e^{-(\lambda + \mu)a} (3 \mu + 2\lambda + \lambda(\lambda + \mu)a) \nonumber\\
&&-\frac{\mu^2}{\lambda^2}e^{-\mu a}(3\mu +2\lambda - \lambda(\lambda+ 2 \mu)a)
\label{eq:fa_barpsi}
\end{IEEEeqnarray}

We also present the conditional complementary cumulative distribution, to describe the probability that the \textit{peak age} surpasses a threshold. In the case of the M/M/1/2* model we have 
\begin{IEEEeqnarray}{rCl}
\mathds{P}(A_k>a)_{M/M/1/2^{*}}&=&\frac{e^{-(\lambda+\mu)a}}{\lambda(\lambda+\mu)(\lambda-\mu)}(\lambda^3-3\mu^3+\lambda\mu(\lambda+\mu)(1+(\lambda-\mu)))\nonumber\\
&&+\frac{e^{-\mu a}}{\lambda(\lambda+\mu)(\lambda-\mu)}(3\mu^3+\lambda(\lambda+\mu)(\lambda-\mu)+\lambda\mu a(\lambda^2+\lambda\mu-2\mu^2))\nonumber\\
&&-\frac{e^{-\lambda a}}{(\lambda+\mu)(\lambda-\mu)}(\lambda^2+\lambda\mu+\mu^2)
\end{IEEEeqnarray}

The \textit{average peak age} can be obtained integrating the complementary cumulative distribution, or summing the expected values for $T_{k-1}$ and $Y_k$. As a final result,
\begin{equation}
\mathds{E}[A_k]_{M/M/1/2^{*}} =\frac{1}{\mu} + \frac{\lambda}{(\lambda+ \mu)^2}+\frac{1}{\lambda}+\frac{1}{\mu}\frac{\lambda}{\lambda+\mu}.
\label{eq:avgAk-mm12}
\end{equation}

\section{Numerical Results}\label{sec:numerical}
In this section, we illustrate the results for \textit{average age} and \textit{peak age} with numerical examples. Unless otherwise stated, we assume unitary service rate ($\mu=1$). When $\mu=1$, and the x-axis has the arrival rate $\lambda$, it can be regarded as the channel utilization $\rho=\lambda/\mu$. In addition to the analytical formulations, some figures present simulation points, obtained from a discrete-event simulator that we have built in MATLAB. The simulation results are presented as circular markers in the curves, and corroborate the analytical results.

\begin{figure}
\centering
\input{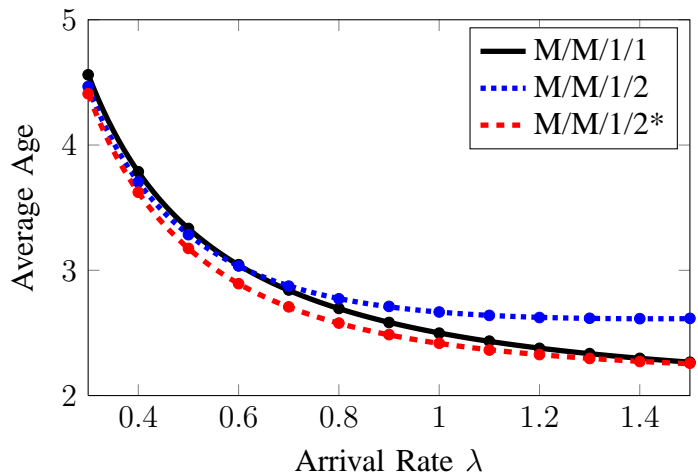}
\caption{\textit{Average age} versus arrival rate ($\lambda$) for the queuing models with packet management. Service rate $\mu=1$.}
\label{fig:AvgArrival} 
\end{figure}

Figure \ref{fig:AvgArrival} illustrates the \textit{average age} for the three queuing models with packet management. The packet replacement scheme yields the smallest \textit{average age}. For example, when $\lambda=0.6$, packet replacement promotes a reduction of \textit{average age} of  approximately $5\%$ in comparison to the other models. The \textit{average age} is decreasing with the channel utilization inside the observed range $0<\rho\leq 1.5$.

\begin{figure}
\centering
\input{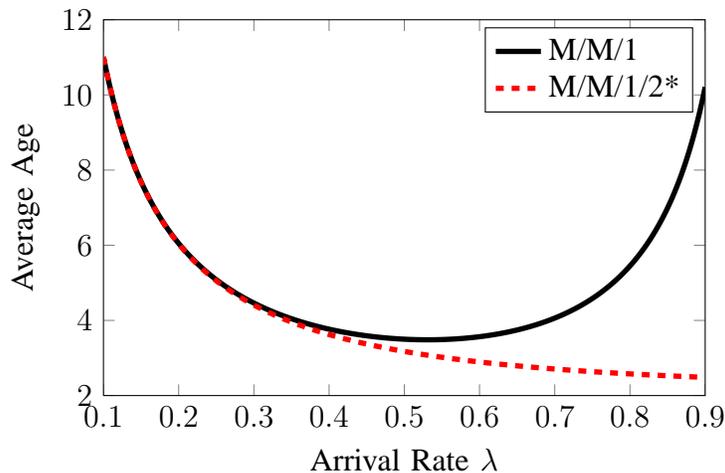}
\caption{\textit{Average age} versus arrival rate ($\lambda$) comparison with model without packet management. Service rate $\mu=1$.}
\label{fig:CompareMM1} 
\end{figure}

Figure \ref{fig:CompareMM1} provides a comparison of the \textit{average age} with packet management modeled as an M/M/1/2* queue and the case without packet management, modeled as an M/M/1 queue, in which every packet that finds the server busy is stored in a buffer for later transmission. The \textit{average age} for the M/M/1 model was presented in \cite{Kaul2012} to be
$$\Delta_{M/M/1}=\frac{1}{\lambda}+\frac{1}{\mu}+\frac{\lambda^2}{\mu-\lambda}.$$
When $\lambda \ll \mu$, the dominant effect is the large interarrival time and the two models have similar behavior because they are idle for a large fraction of time. As the arrival rate $\lambda$ increases, the dominant effect in the M/M/1 system is the queuing time which approaches infinity as the arrival rate approaches the service rate. As a result, the \textit{average age} increases to infinity as the arrival rate approaches the service rate. This effect is eliminated with the packet management scheme, which promotes significant improvement in the \textit{average age}, particularly for $\rho>0.5$. 

\begin{figure}
\centering
\input{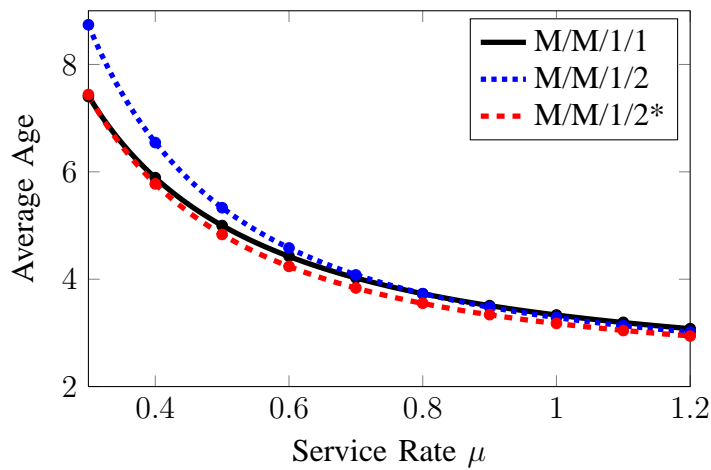}
\caption{\textit{Average age} versus service rate ($\mu$) for the queuing models with packet management. Arrival rate $\lambda=0.5$.}
\label{fig:avg_departure} 
\end{figure}

Figure \ref{fig:avg_departure} shows the \textit{average age} versus the service rate, with fixed arrival rate $\lambda=0.5$. In all cases the M/M/1/2* promotes the smallest \textit{average age}, but the schemes have a common limit value as $\mu$ goes to infinity, which is $1/\lambda$. That is, for very small service times, the average age is limited by the interarrival times. 

\begin{figure}
\centering
\begin{tikzpicture}
\node at (0,0){\input{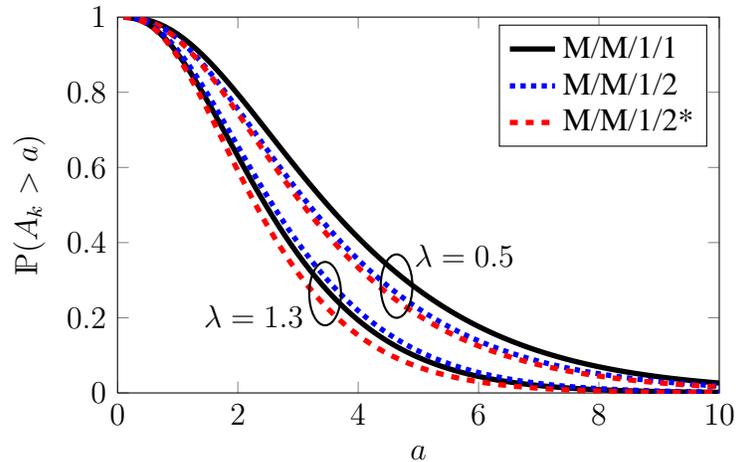}};
\draw [thick] (0.3,-0.7) ellipse (6pt and 12pt);
\node at (1.2,-0.3){$\lambda=0.5$};
\draw [thick] (-0.65,-0.8) ellipse (6pt and 12pt);
\node at (-1.6,-1.1){$\lambda=1.3$};
\end{tikzpicture}
\caption{\textit{Peak age} distribution: $\mathds{P}(A_k>a)$ versus threshold $a$. Service rate $\mu=1$. Arrival rates $\lambda=0.5$ and $\lambda=1.3$.}
\label{fig:CCDF_peak} 
\end{figure}

The complementary cumulative distribution function for the \textit{peak age} is presented in Figure \ref{fig:CCDF_peak}. We consider a service rate $\mu=1$, with arrival rates $\lambda=0.5$ and $\lambda=1.3$. The probability that the \textit{peak age} is larger than a given threshold $a$ can be reduced with larger arrival rates for all three schemes. Comparing the two groups of curves, we also note that the M/M/1/2* model produces the best \textit{peak age} results for arrival rates below and above the service rate. The M/M/1/2 model presents the second best results for small arrival rates, but for large arrival rates the M/M/1/1 model performs better than the M/M/1/2 with respect to the \textit{peak age}.   

\begin{figure}
\centering
\input{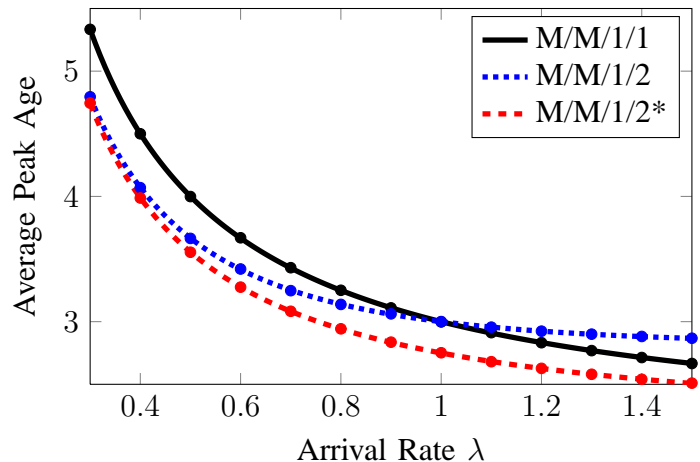}
\caption{\textit{Average peak age} versus arrival rate ($\lambda$) for the queuing models with packet management. Service rate $\mu=1$.}
\label{fig:peak_arrival} 
\end{figure}

Figure \ref{fig:peak_arrival} illustrates the \textit{average peak age} for the three queuing models. It corroborates the conclusion that keeping a packet in the buffer is preferable in the case of small arrival rates, while discarding all the packets that find the server busy could be adopted for very large arrival rates ($\lambda>\mu$). The M/M/1/2* model with packet replacement presents the best results, and is the most adequate model for applications that require the \textit{age of information} available to the receiver to be below a certain threshold. That is the case when the outdated information looses its value due to small correlation with the current state of the process under observation. 

\section{Final Remarks}\label{sec:conclusion}

The concept of \textit{age of information} is relevant to any system concerned with the timeliness of information. This is the case when an action is taken based on available information, but the information loses its value with time. The characterization of \textit{age of information} is still incipient, and the investigation of simple models remains important to understand  its impact on the performance of communication systems.

This work provides contributions to the initial steps in the characterization of \textit{age} in communication systems that transmit status update messages. We have considered that the source node can manage the samples of a process of interest, deciding to discard or to transmit status updates to the destination. We observe that package management with packet replacement in the buffer promotes smaller \textit{average age}, when compared to schemes without replacement. The proposed scheme is asymptotically optimal, achieving a lower bound for \textit{average age} while avoiding the waste of network resources in the transmission of stale information.

In addition to the \textit{average age}, we proposed a new metric, called \textit{peak age}. The \textit{peak age} is a suitable metric to characterize the \textit{age of information} in applications that impose a threshold on the value of \textit{age}. The \textit{peak age} has the advantage of a more simple mathematical formulation, which will certainly benefit future investigations regarding the optimization of a network with respect to the timeliness of the transmitted information.

\bibliographystyle{IEEEtran}
\bibliography{IEEEabrv,EnergyEfficiency_Age}

\end{document}